\documentclass[a4paper, 12pt, openany, oneside]{article}
\usepackage[cp1251]{inputenc} 
\usepackage[english, russian]{babel}
\usepackage{amsmath, latexsym, amssymb, array, graphics, amsfonts, bm}
\usepackage{indentfirst}
\renewcommand{\baselinestretch}{1.2}
\usepackage[dvips]{graphicx}
\usepackage[flushleft,small]{caption2}

\begin{document}

\thispagestyle{empty}
\large
\renewcommand{\refname}{\begin{center} REFERENCES\end{center}}
\renewcommand{\abstractname}{\begin{center} Abstract\end{center}}
\renewcommand{\contentsname}{\begin{center} Contents\end{center}}
 \begin{center}
\bf A new method for solving of vector problems for kinetic equations
with Maxwell boundary conditions
\end{center}\medskip
\begin{center}
  \bf
  A. V. Latyshev\footnote{$avlatyshev@mail.ru$}
\end{center}\medskip

\begin{center}
{\it Faculty of Physics and Mathematics,\\ Moscow State Regional
University, 105005,\\ Moscow, Radio str., 10--A}
\end{center}\medskip

\tableofcontents
\setcounter{secnumdepth}{4}
\bigskip
\bigskip
\begin{abstract}
In the present work the classical problem of the kinetic theory of gases
(the Smoluchowsky' problem about temperature jump in rarefied gas)
is considered. The rarefied gas fills half-space
over a flat firm surface. logarithmic gradient of temperature  is set
far from surface.
The kinetic equation  with
modelling integral of collisions in the form of BGK-model
(Bhatnagar, Gross and Krook) is used.

The general  mirror-diffuse boundary conditions
of molecules reflexions of gas from a wall on  border of half-space
(Maxwell conditions) are considered.
Expanding distribution function on two
orthogonal directions in space of velocities, the
Smolu\-chow\-sky' problem to the solution of the
homogeneous vector one-dimensional
and one-velocity kinetic equation with a matrix kernel is reduced.

Then generalization of  source-method is used and boundary conditions
include in non-homogeneous vector kinetic equation.
The solution in the form of Fourier integral is searched.
The problem is reduced to
the solution of vector Fredholm integral equation of the second
sort with  matrix kernel.

The solution of Fredholm equation in the form of Neumann's polynoms with
vector coefficients is searched.
The system vector algebraic interengaged equations turns out.
The solution of this system is under construction
in the form of Neumann's polynoms.
Comparison with well-known Barichello---Siewert'  high-exact results
is made. Zero and the first approach of jumps of temperature
and numerical density are received.
It is shown, that transition from the zero
to the first approach raises 10 times accuracy in calculation
coefficients of temperature and concentration jump.
\medskip

\medskip

transport theory of gases.
\end{abstract}

\begin{center}
\item{} \section{\bf Introduction}
\end{center}

The problem about temperature jump is known from the end of XIX century
\cite{1}. M.Smoluhovsky has constructed in \cite{1} the theory
of temperature jump in the rarefied gas. Since then this problem
the invariable attention already draws for a long time to itself
(history of this question see in \cite{2} and \cite{3}).

This problem has been solved  with use of approximate and
numerical methods as for the modelling equations, and
for full (non-linear)  Boltzmann equation \cite{4} - \cite{18}.

In 1972 in work \cite{19} the problem about
temperature jump with use of the modelling Boltzmann equation
with collisional integral BGK (Bhatnagar, Gross, Krook)
and with frequency of collisions of the molecules, proportional to the module
of velocities of molecules has analytically been solved.

Attempts of the exact solution of this problem
about temperature jump with diffusion boundary conditions and
with use of the modelling Boltzmann equation
with collisional integral  BGK with constant frequency
of collisions of molecules \cite{20} - \cite{23}
begin with this moment.

The solution of this problem encounters considerable difficulties.
This problem is formulated in the form of a vector boundary problem.
The solution of last problem meets the solution of a vector
boundary value
Riemann---Hilbert problem with the matrix coefficient, having points
of branchings.
These difficulties have been overcome only in 1990 in our work
\cite{24} where the analytical solution of the Smoluchowsky
problem has been received.

By the method developed in \cite{24}, further problems
about temperature jump in molecular gases \cite{25}, and also
in metal \cite{26} have been  analytically solved.

Along with the  Smoluchowsky' problem the big interest represents studying
of behaviour of gas at weak evaporation (condensation) from a surface.
These problems are called as the generalized  Smoluchowsky' problem
in view of that boundary conditions in these problems differ slightly.

Let us notice, that in works \cite{17,18} the various
kinetic models were used, in particular, the model of Shakhov
(or, S-model, see \cite{27}).

For the solution of boundary half-space problems with accommodation
some methods \cite{28} - \cite{30} have been developed, allowing to receive
the solution of this problem with any degree of accuracy.

In the present work the generalized source-method from \cite{30}
extends on a vector case to which the problem
about temperature jump is reduced.
Thus the effective method of the solution of
boundary problems with mirror - diffusion boundary conditions
(Maxwell conditions) develops. We will notice, that the method from
\cite{30} has already been applied in
problems of electro\-dy\-namics of plasma \cite{31} and in condensate problems
of Bose---Einstein.

At the heart of an offered method the idea lays to include the boundary
condition  in the form of a source in the kinetic equation.

The method basis consists in the following. At first
in half-space $x> 0$ are formulated a vector problem about the temperature
jump with boundary Maxwell conditions. Then unknown function
continuations in conjugated half-space $x <0$ in the even method on
spatial and on velocity variables. In half-space
$x <0$ also are formulated the problem about temperature jump.

Now let us expand unknown function (which we will name also
distribution function) on two composed: Chapman---Enskog'
distribution function $h_{as}(x,\mu)$ and the second part of function
distributions $h_c(x,\mu)$, correspoding to conti\-nu\-ous spectrum
(see \cite{30} )
$$
h(x,\mu)=h_{as}(x,\mu)+h_c(x,\mu)
$$
($as \equiv asymptotic, c\equiv continuous$).

Owing to that Chapman---Enskog' distribution function
there is a linear combination of discrete solutions of the initial equation,
function $h_c(x,\mu)$ also is the solution of the kinetic
equations. Function $h_c(x,\mu)$ vanishes in zero far from
wall. On a wall this function satisfies to boundary Maxwell condition.

Further we will transform the equation for function
$h_c(x,\mu)$. We include in this equation boundary condition on
wall for function $h_c(x,\mu) $ in the form of a member
of source-type laying in a plane $x=0$.

We will underline, that function
$h_c(x,\mu)$ satisfies to the received equation in
both conjugated half-spaces $x <0$ and $x> 0$.

We solve this equation in the second and the fourth
quarters of a phase plane $(x,\mu)$ as the linear
differential equation of the first order, considering known
the right part of the equation $U_c(x)$.
From the received solutions we  found the boundary
values of unknown function $h^{\pm}(x,\mu)$ at $x =\pm 0$,
entering into the equation.

Now we expand  by Fourier integrals unk\-nown function $h_c(x,\mu)$,
an unknown right part $U_c(x)$ and Dirac delta-function.
Boundary values of the unknown functions $h_c^{\pm}(0,\mu)$
are thus expressed by the same integral on the spectral
density $E(k)$ functions $U_c(x)$.

Substitution of Fourier integrals in the kinetic equation and
expression for the right part $U_c(x)$ leads to the vector characteristic
system of equations. If to exclude from this system the spectral
density $\Phi(k,\mu)$ of function $h_c(x,\mu) $, we will receive
vector Fredholm integral equation of the second sort.

Believing the gradient of the logarithm of temperature is setting,
we will expand the unknown quantities of temperature and
concentration jumps and also spectral density
by polynoms on degrees of coefficient of
diffusion $q$ (these are Neumann's polynoms). On this way we receive
system of the hooked equations on coefficients of polynoms for
spectral density. Thus all equations on coefficients of
spectral density have singularity (a pole
of the second order in zero). Excepting these singularities
consistently, we will construct all members of the polynoms
for quantities of temperature and
concentration jumps and for spectral density $E(k)$.

\begin{center}
 \item{}\section{Statement problem}
\end{center}

Let the rarefied one-nuclear gas occupies half-space $x>0$
over the flat firm surface laying in a plane $x=0$.
Far from a wall the logarithmic gradient of temperature is set
$$
g_T=\Big(\dfrac{d \ln T}{d x}\Big)_{x=+\infty}.
$$

We take the stationary kinetic equation of relaxation type with
collisional integral BGK (Bhatnagar, Gross and Krook)
$$
v_x\dfrac{\partial f(x,{\bf v})}{\partial x}=
\dfrac{f_{eq}(x,v)-f(x,{\bf v})}{\tau},
$$
where  $\tau$ is the time between two consecutive collisions
of molecules, $\nu=1/\tau$ is the collisional frequence of gaseous
molecules, $f_{eq}$ is the equilibrium distribution function,
$$
f_{eq}(x,v)=n(x)\Big(\dfrac{m}{2\pi k T}\Big)^{3/2}\exp
\Big(-\dfrac{m}{2kT(x)}v^2\Big),
$$
where  $m$ is the mass of molecule, $k$ is the Boltzmann constant,
$T(x)$ is the gas temperature,
$$
T(x)=\dfrac{2E(x)}{3kn(x)},\qquad E(x)=\int \dfrac{m}{2}v^2f(x,{\bf v})d^3v,
$$
$n(x)$ is the gas number density (concentration),
$$
n(x)=\int f(x,{\bf v})d^3v.
$$

Further we will be linearize the kinetic equation and  search
distribution function in the form
$$
f(x,\mathbf{v})=f_0(v)(1+\varphi(x,\mathbf{v})),
\eqno{(2.1)}
$$
where $f_0(v)$ is the absolute Maxwellian,
$$
f_0(v)=n_0\Big(\dfrac{m}{2\pi kT_0}\Big)^{3/2}
\exp\Big(-\dfrac{mv^2}{2kT_0}\Big),
$$
where $n_0, T_0$ are number density and gas temperature in some point,
for example, in origin of coordinates.

Let us be linerize distribution of numerical density and
temperature concerning parametres $n_0$ and $T_0$
$$
n(x)=n_0+\delta n(x), \qquad T(x)=T_0+\delta T(x).
$$

According to (2.1) for distribution of numerical density we have
$$
n(x)=\int f_0(v)[1+\varphi(x,{\bf v})]d^3v=n_0+\delta n(x),
$$
where
$$
n_0=\int f_0(v)d^3v, \qquad \delta n(x)=\int f_0(v)\varphi(x,{\bf v})d^3v.
$$

For distribution of temperature we receive
$$
T(x)=\dfrac{2}{3kn(x)}\int \dfrac{mv^2}{2}f_0(v)[1+\varphi(x,{\bf v})]d^3v.
$$
We notice that
$$
\dfrac{n_0}{n(x)}=1-\dfrac{\delta n}{n_0}+o(h),\qquad h\to 0.
$$

Now for temperature we receive
$$
T(x)=T_0\dfrac{2}{3n_0}\Big(1-\dfrac{\delta n}{n_0}\Big)\int
\dfrac{mv^2}{2kT_0}f_0(v)[1+\varphi(x,{\bf v})]d^3v=
$$
$$
=\dfrac{2T_0}{3n_0}\Big(1-\dfrac{\delta n}{n_0}\Big)\int
\dfrac{mv^2}{2kT_0}f_0(v)d^3v+
\dfrac{2T_0}{3n_0}\Big(1-\dfrac{\delta n}{n_0}\Big)\int
\dfrac{mv^2}{2kT_0}f_0(v)\varphi(x,{\bf v})d^3v.
$$

We notice that
$$
\dfrac{2}{3n_0}\int\dfrac{mv^2}{2kT_0}f_0(v)d^3v=1.
$$

Hence, for relative change of temperature it is had
$$
\dfrac{\delta T}{T_0}=-\dfrac{\delta n}{n_0}+\dfrac{2}{3n_0}\int
\dfrac{mv^2}{2kT_0}f_0(v)\varphi(x,{\bf v})d^3v.
$$

We will be linearize equilibrium function of distribution
$$
f_{eq}(v)=f_0(v)\Big[1+\dfrac{\delta n(x)}{n_0}+\Big(\dfrac{mv^2}{2kT_0}-
\dfrac{3}{2}\Big)\dfrac{\delta T(x)}{T_0}\Big].
$$

We receive the following equation after linearizing
the kinetic BGK--equation according to (2.1)
$$
v_x\dfrac{\partial \varphi(x,{\bf v})}{ \partial x}=
\nu \Big[\dfrac{\delta n(x)}{n_0}+\Big(\dfrac{mv^2}{2kT_0}-
\dfrac{3}{2}\Big)\dfrac{\delta T(x)}{T_0}-\varphi(x,{\bf v})\Big].
$$

Let us enter dimensionless velocities and parametres ---
dimensionless velocity of molecules
$\mathbf{C}=\sqrt{\beta}\mathbf{v}=\dfrac{\mathbf{v}}{v_T}$,
where
$\beta=\dfrac{m}{2kT_0}$,
dimensionless time $t_1=\nu t$, dimensionless coordinate
$$
x_1=\nu\sqrt{\dfrac{2kT_0}{m}}x=\dfrac{x}{v_T\tau}=\dfrac{x}{l},
$$
where $l=v_T\tau$ is the mean free path of gaseous molecules,
$v_T=\dfrac{1}{\sqrt{\beta}}=\sqrt{\dfrac{2kT_0}{m}}$
is the thermal velocity of the molecules movements,
having an order of velocity of a sound.

Now the kinetic equation will be transformed to the form
$$
C_x\dfrac{\partial \varphi}{\partial x_1}+\varphi(x_1,{\bf C})=
\dfrac{\delta n(x_1)}{n_0}
+\Big(C^2-\dfrac{3}{2}\Big)\dfrac{\delta T(x_1)}{T_0}.
$$

Here
$$
\dfrac{\delta n(x_1)}{n_0}=\dfrac{1}{\pi^{3/2}}
\int e^{-C^2}\varphi(x_1,{\bf C})d^3C,
$$
$$
\dfrac{\delta T(x_1)}{T_0}=\dfrac{2}{3\pi^{3/2}}
\int e^{-C^2}\Big(C^2-\dfrac{3}{2}\Big)\varphi(x_1,{\bf C})d^3C.
$$

Further a variable $x_1$ we will designate again through $x$.

Let us transform the  linear kinetic equation to the form
$$
C_x\dfrac{\partial \varphi}{\partial x}+\varphi(x,\mathbf{v})=
\dfrac{1}{\pi^{3/2}}\int K({\bf C},{\bf C'})\varphi(x,{\bf C})
e^{-C'^2}d^3C'
\eqno{(2.2)}
$$
with kernel
$$
K({\bf C},{\bf C'})=1+\dfrac{2}{3}\Big(C^2-\dfrac{3}{2}\Big)
\Big(C'^2-\dfrac{3}{2}\Big).
$$

It is easy to check up, that the equation (2.2) has the following
partial solutions
$$
\begin{array}{l}
  \varphi_1(x,\mu)=1, \\
  \varphi_2(x,\mu)=C^2-\dfrac{3}{2}, \\
  \varphi_3(x,\mu)=(x-C_x)\Big(C^2-\dfrac{5}{2}\Big).
\end{array}
$$

Let us construct asymptotic  Chapman---Enskog distribution in
the form of the linear combination of partial solutions of the equation (2.2)
with arbitrary constants
$$
\varphi_{as}(x,\mu)=A_0+A_1\Big(C^2-\dfrac{3}{2}\Big)+A_2(x-\mu)\Big(C^2-
\dfrac{5}{2}\Big),
\eqno{(2.3)}
$$
where $A_0,A_1,A_2$ are arbitrary constants.

For finding of these constants we will take advantage of definition
of the macroscopical parametres. From definition of numerical
density (concentration)
$$
n(x)=\int f(x, \mathbf{v})\,d^3v
$$
follows, that the extrapolated concentration of gas on the wall is equal
$$
n_e=n_{as}(0)=\int f_{as}(0, \mathbf{v})\,d^3v=
\int f_0(v)(1+\varphi_{as}(0, \mathbf{v}))\,d^3v=
$$
$$
=n_0( \beta/ \pi)^{3/2}\int \exp(- C^2)(1+\varphi_{as}(0, \mathbf{ C}))\,d^3v.
$$

From here we have
$$
\dfrac{n_e}{n_0}= \pi^{-3/2}\int
\exp(- C^2)(1+\varphi_{as}(0, \mathbf{ C}))\,d^3 C,
$$
or
$$
\dfrac{n_e}{n_0}=1+ \pi^{-3/2}
\int \exp(- C^2)\varphi_{as}(0, \mathbf{ C})\,d^3 C.
$$

Hence, the quantity of jump of concentration is searched  under the formula
$$
\dfrac{n_e-n_0}{n_0}= \varepsilon_n= \pi^{-3/2}
\int \exp(- C^2)\varphi_{as}(0, \mathbf{C})\,d^3 C.
$$

Substituting expression (2.3) in this equality, we have
$$
\varepsilon_n=A_0.
$$

Setting gradient of temperature far from a wall means, that
temperature distribution in half-space looks like
$$
T(x)=T_e+ \left( \dfrac{dT}{dx}\right)_{x=+\infty} x=T_e+G_Tx,\quad
\text{где}\quad G_T=\left( \dfrac{dT}{dx}\right)_{x=+ \infty}.
$$

This distribution we will present in the form
$$
T(x)=T_0\Big( \dfrac{T_e}{T_0}+g_T x\Big)=
T_0\left(1+ \dfrac{T_e-T_0}{T_0}+ g_T x \right), \quad x \to + \infty,
$$
or
$$
T(x)=T_0(1+ \varepsilon_T+ g_Tx), \qquad x \to + \infty,
\eqno{(2.4)}
$$
where
$$
\varepsilon_T=\dfrac{T_e-T_0}{T_0}
$$
is the required quantity of temperature jump.

From expression (2.4) it is visible, that relative change of
temperature far from walls it is described by linear function
$$
\dfrac{ \delta T_{as}(x)}{T_0}= \dfrac{T(x)-T_0}{T_0}=
\varepsilon_T+g_T x, \quad x \to + \infty.
\eqno{(2.5)}
$$

Relative change of temperature we will present in the form
$$
 \dfrac{ \delta T(x)}{T_0}= \dfrac{2}{3} \pi^{-3/2}
 \int \exp(- C^2)( C^2-\dfrac{3}{2})\varphi(x, \mathbf{ C} )\,d^3 C.
$$

Far from a wall relative change of temperature
transforms as follows
$$
 \dfrac{ \delta T_{as}(x)}{T_0}= \dfrac{2}{3} \pi^{-3/2}
 \int \exp(- C^2)( C^2-\dfrac{3}{2})\varphi_{as}(x, \mathbf{ C} )\,d^3 C.
$$

Substituting in this equality expression (2.3) for
$\varphi_{as}(x, \mathbf{ C})$, we find, that
$$
\dfrac{ \delta T(x)}{T_0}=A_2+A_3 x \quad (x\to + \infty).
\eqno{(2.6)}
$$

Comparing expressions (2.5) and (2.6), we find, that
$
A_2= \varepsilon_T, \qquad A_3=g_T.
$

Thus, asymptotic part of function of distribution
(at $x \to + \infty $) it is constructed and on the basis
stated above transforms in the form
$$
\varphi_{as}(x, \mathbf{ C})= \varepsilon_n+\varepsilon_T( C^2-
\dfrac{3}{2})+g_T(x- C_x)( C^2- \dfrac{5}{2}).
$$

Let us formulate down boundary conditions to the equation (2.2). At first
let us formulate mirror--diffusion boundary condition on a wall
for full function of distribution
$$
f(+0,{\bf v})=qf_0(v)+(1-q)f(+0,-v_x,v_y,v_z), \qquad v_x>0.
$$

Here $q $ is the accommodation coefficient, i.e. a part of the molecules flying
after reflexion from a wall with Maxwell distribution on
velocities, $1-q$ is the part of the molecules reflected from a wall purely
mirror.

Using (2.1), from here we receive a boundary condition of a problem onto
wall
$$
\varphi(0,{\bf C})=(1-q)\varphi(x,-C_x,C_y,C_z), \qquad C_x>0.
\eqno{(2.7)}
$$

Let us demand, that far from a wall distribution function
passed into Chapman---Enskog distribution
with coordinate growth
$$
f(x,{\bf v})=f_0(v)\Bigg[1+\varepsilon_n+
\varepsilon_T\Big(\dfrac{mv^2}{2kT_0}-\dfrac{3}{2}\Big)+
g_T\Big(x-\sqrt{\dfrac{m}{2kT_0}}v_x\Big)
\Big(\dfrac{mv^2}{2kT_0}-\dfrac{5}{2}\Big)\Bigg],\; x\to +\infty.
$$

From here according to (2.1) for function $ \varphi $ we receive
the following boundary conditions
$$
\varphi(x,\mu)=\varphi_{as}(x,\mu)+o(1), \qquad x\to+\infty.
\eqno{(2.8)}
$$

Here $\varphi_{as}(x,\mu)$ is the asymptotic Chapman---Enskog
distribution, entered above.

So, the boundary problem about finding of jumps of temperature
and concentration of gas
(vapor) over a flat surface consists in finding of the such
solution of the equation (2.2),
which satisfies to boundary conditions (2.7) and (2.8).

\begin{center}
 \item{} \section{Reduction to vector boundary problem}
\end{center}

If to use substitution
$$
\varphi(x,{\bf C})=h_1(x,\mu)+\Big(C^2-\dfrac{3}{2}\Big)h_2(x,\mu),\qquad
\mu=C_x,
\eqno{(3.1)}
$$
that equation (2.2) is reduced breaks up to two equations
$$
\mu\dfrac{\partial h_1}{\partial x}+h_1(x,\mu)=$$$$=
\dfrac{1}{\sqrt{\pi}}\int\limits_{-\infty}^{\infty}e^{-\mu'^2}
\Big[h_1(x,\mu')+\Big(\mu'^2-\dfrac{1}{2}\Big)h_2(x,\mu')\Big]d\mu'
$$
and
$$
\mu\dfrac{\partial h_2}{\partial x}+h_2(x,\mu)=$$$$=
\dfrac{2}{3\sqrt{\pi}}\int\limits_{-\infty}^{\infty}e^{-\mu'^2}
\Big[\Big(\mu'^2-\dfrac{1}{2}\Big)h_1(x,\mu')+
\Big(\mu'^4-\mu'^2+\dfrac{5}{4}\Big)h_2(x,\mu')\Big]d\mu'.
$$

This equation we will present in the vector form
$$
\mu\dfrac{\partial h}{\partial x}+h(x,\mu)=\dfrac{1}{\sqrt{\pi}}
\int\limits_{-\infty}^{\infty}e^{-\mu'^2}K(\mu')h(x,\mu')d\mu'.
\eqno{(3.2)}
$$

Here $h(x,\mu)$ is the vector-column
$$
h(x,\mu)=\left(
           \begin{array}{c}
             h_1(x,\mu) \\
             h_2(x,\mu) \\
           \end{array}
         \right),
$$
and matrix kernel have the following form
$$\extrarowheight=16pt
K(\mu)=\left(
         \begin{array}{cc}
           1 & \Big(\mu^2-\dfrac{1}{2}\Big) \\
          \dfrac{2}{3}\Big(\mu^2-\dfrac{1}{2}\Big) & \dfrac{2}{3}\Big[
           \Big(\mu^2-\dfrac{1}{2}\Big)^2+1\Big] \\
         \end{array}
       \right).
$$

The right part of the equation (3.2)
$$
U(x)=\dfrac{1}{\sqrt{\pi}}
\int\limits_{-\infty}^{\infty}e^{-\mu'^2}K(\mu')h(x,\mu')d\mu'
$$
has clear physical sense. The vector-column $U (x) $ looks like
$$\extrarowheight=10pt
U(x)=\left(
             \begin{array}{c}
               \dfrac{\delta n(x)}{n_0} \\
               \dfrac{\delta T(x)}{T_0} \\
             \end{array}
           \right),
$$
i.e., components of this vector consist of the relative
changes of numerical density of gas and relative change
temperatures (concerning equilibrium values $ (n_0, T_0) $).
It is possible to present this vector in the form
$$
U(x)=U_{as}(x)+U_{c}(x),
$$
where
$$\extrarowheight=10pt
U_{as}(x)=\left(
             \begin{array}{c}
               \dfrac{\delta n_{as}(x)}{n_0} \\
               \dfrac{\delta T_{as}(x)}{T_0} \\
             \end{array}
           \right),\qquad
U_{c}(x)=\left(
             \begin{array}{c}
               \dfrac{\delta n_{c}(x)}{n_0} \\
               \dfrac{\delta T_{c}(x)}{T_0} \\
             \end{array}
           \right).
$$

According to (3.1) from (2.7) and (2.8) for the vector-functions $h(x,\mu)$
we receive the following vector boundary conditions
$$
h(+0,\mu)=(1-q)h(+0,-\mu), \qquad \mu>0,
\eqno{(3.3)}
$$
and
$$
h(x,\mu)=h_{as}(x,\mu)+o(1),\qquad x\to +\infty,
\eqno{(3.4)}
$$
where
$$
h_{as}(x,\mu)=\left(
                \begin{array}{c}
                  \varepsilon_n \\
                  \varepsilon_T \\
                \end{array}
              \right)+g_T(x-\mu)\left(
                \begin{array}{r}
                  -1\\
                  1\\
                \end{array}
              \right).
$$

Function $h_{as}(x,\mu)$ is the solution of the equation (3.2).
Hence, if to search for the solution of the equation (3.2) in the
form
$$
h(x,\mu)=h_{as}(x,\mu)+h_c(x,\mu),
\eqno{(3.5)}
$$
then function $h_{c}(x,\mu)$ is still searched from the equation (3.2)
$$
\mu\dfrac{\partial h_c}{\partial x}+h_c(x,\mu)=\dfrac{1}{\sqrt{\pi}}
\int\limits_{-\infty}^{\infty}e^{-\mu'^2}K(\mu')h_c(x,\mu')d\mu'.
\eqno{(3.6)}
$$
and  boundary conditions (3.3) and (3.4) will be transformed thanking (3.5)
to the following form
$$
h_c(+0,\mu)=h_0^+(\mu)+(1-q)h_c(+0,-\mu), \qquad \mu>0,
\eqno{(3.7)}
$$
$$
h_c(+\infty,\mu)={\bf 0}, \qquad \quad
{\bf 0}=\left(
          \begin{array}{c}
            0 \\
            0 \\
          \end{array}
        \right).
\eqno{(3.8)}
$$

Here
$$
h_0^+(\mu)=-q\left(\begin{array}{c}
                  \varepsilon_n \\
                  \varepsilon_T \\
                \end{array}\right)+(2-q)g_T^+\mu\left(
                \begin{array}{r}
                  -1\\
                 1\\
                \end{array}
              \right).
\eqno{(3.9)}
$$

Let us solve further the problem consisting of the solution
of the equation (3.6) with boundary conditions (3.7) -- (3.10).

\section{Kinetic equation with  source}

For solution of this problem the auxiliary problem is required to us in
"negative" \, half-space $x <0$. That it
to formulate, we will continue function $h(x,\mu)$ as follows
$$
h(x,\mu)=h(-x,-\mu).
\eqno{(4.1)}
$$

Let us notice, that at continuation (4.1) logarithmic
temperature gradient $g_T $,
which for "positive" \, half-spaces we will designate through
$g_T^+$, changes the sign
$$
g_T^-=\Big(\dfrac{d\ln T }{dx}\Big)_{x=-\infty}=
-\Big(\dfrac{d\ln T }{dx}\Big)_{x=+\infty}=-g_T^+.
$$

Besides, we will notice, that function $h_{as}(+0,\mu)$ automatically
satisfies to equality (4.1):
$h_{as}(+0,\mu)=h_{as}(-0,-\mu)$. This equality means, that
the equality (3.1) is carried out for function $h_{as}(x,\mu)$:
$h_{as}(x,\mu)=h_{as}(-x,-\mu)$.

Hence, boundary conditions in "negative" \,
space are formulated as follows
$$
h_c(-0,\mu)=h_0^-(\mu)+(1-q)h_c(-0,-\mu), \qquad \mu<0,
$$
$$
h_c(-\infty,\mu)={\bf 0}.
$$

Here
$$
h_0^-(\mu)=-q\left(\begin{array}{c}
                  \varepsilon_n \\
                  \varepsilon_T \\
                \end{array}\right)+(2-q)g_T^-\mu\left(
                \begin{array}{r}
                  -1\\
                  1\\
                \end{array}
              \right).
$$

Let us unite both problems --- in "positive" \, and "negative" \,
half-spaces --- in one, having included boundary conditions in
the kinetic equation by means of member of type of the source
$$
\mu\dfrac{\partial h_c}{\partial x}+h_c(x,\mu)=U_c(x)+|\mu|\delta(x)
\Big[h_0^{\pm}(\mu)-qh_c(\mp 0, \mu)\Big].
\eqno{(4.2)}
$$

Here
$$
U_c(x)=\dfrac{1}{\sqrt{\pi}}\int\limits_{-\infty}^{\infty}
e^{-\mu^2}K(\mu)h_c(x,\mu)d\mu,
\eqno{(4.3)}
$$
$$
h_0^{\pm}(\mu)=-q\left(\begin{array}{c}
                  \varepsilon_n \\
                  \varepsilon_T \\
                \end{array}\right)+(2-q)g_T^+|\mu|\left(
                \begin{array}{r}
                  -1\\
                 1\\
                \end{array}
              \right),
$$

$$
h_c(\mp 0,\mu)=\lim\limits_{x\to \mp 0,\pm x<0}h_c(x,\mu),\qquad
\pm \mu>0.
$$

These function $h_c(\mp 0,\mu)$  are finding from equalities
$$
h_c^+(x,\mu)=-\dfrac{1}{\mu}\int\limits_{x}^{+\infty}e^{t/\mu}U_c(t)dt,\qquad
h_c^-(x,\mu)=\dfrac{1}{\mu}\int\limits_{-\infty}^{x}e^{t/\mu}U_c(t)dt.
\eqno{(4.4)}
$$

\begin{center}
  \item{}\section{Vector Fredholm equation  of second sort}
\end{center}

The solution of the equations (4.2) and (4.3) we
search in the form of Fourier integrals
$$
U_c(x)=\dfrac{1}{2\pi}\int\limits_{-\infty}^{\infty}e^{ikx}E(k)dk,\;\qquad
\delta(x)=\dfrac{1}{2\pi}\int\limits_{-\infty}^{\infty}e^{ikx}dk,\;
\eqno{(5.1)}
$$
$$
h_c(x,\mu)=\dfrac{1}{2\pi}\int\limits_{-\infty}^{\infty}
e^{ikx}\Phi(k,\mu)dk.
\eqno{(5.2)}
$$

From equalities (4.3) and (5.1) follows, that
$$
E(k)=\dfrac{1}{\sqrt{\pi}}\int\limits_{-\infty}^{\infty}e^{-\mu^2}K(\mu)
\Phi(k,\mu)dk.
\eqno{(5.3)}
$$

Two following equalities follow from equalities (4.4)
$$
h_c^{\pm}(0,\mu)=\dfrac{1}{2\pi}\int\limits_{-\infty}^{\infty}
\dfrac{E(k_1)dk_1}{1+ik_1\mu}=\dfrac{1}{\pi}
\int\limits_{0}^{\infty}\dfrac{E(k_1)dk_1}{1+k_1^2\mu^2}.
\eqno{(5.4)}
$$

From the kinetic equation (4.2) by means of (5.4) it is found
$$
\Phi(k,\mu)=\dfrac{E(k)}{1+ik\mu}-q
\left(\begin{array}{c}\varepsilon_n \\\varepsilon_T \\
\end{array}\right)\dfrac{|\mu|}{1+ik\mu}+
$$
$$
+(2-q)g_T \left(\begin{array}{r}-1 \\
1\\\end{array}\right)\dfrac{\mu^2}{1+ik\mu}-
\dfrac{|\mu|}{1+ik\mu}\dfrac{q}{\sqrt{\pi}}\int\limits_{0}^{\infty}
\dfrac{E(k_1)dk_1}{1+k_1^2\mu^2}.
\eqno{(5.5)}
$$

Substituting (5.5) in (5.3), we come to the vector
integral Fredholm equation  of the second sort
$$
L(k)E(k)=-q\hat T_1(k)
\left(\begin{array}{c}\varepsilon_n \\\varepsilon_T \\
\end{array}\right)+(2-q)g_T\hat T_2(k)\left(\begin{array}{r}-1 \\
1\\\end{array}\right)-
\dfrac{q}{\pi}
\int\limits_{0}^{\infty}\hat J(k,k_1)E(k_1)dk_1.
\eqno{(5.6)}
$$

Here $L(k)$ is the dispersion matrix-function
$$
L(k)=E_2-\dfrac{1}{\sqrt{\pi}}\int\limits_{-\infty}^{\infty}
\dfrac{e^{-\mu^2}K(\mu)d\mu}{1+ik\mu}=$$$$=
E_2-
\dfrac{2}{\sqrt{\pi}}\int\limits_{0}^{\infty}
\dfrac{e^{-\mu^2}K(\mu)d\mu}{1+k^2\mu^2}=E_2-\hat T_0(k),
$$
where $E_2$ is the unit matrix of the second order,
$$
\hat  T_n(k)=\dfrac{2}{\sqrt{\pi}}\int\limits_{0}^{\infty}
\dfrac{e^{-\mu^2}K(\mu)\mu^nd\mu}{1+k^2\mu^2},\qquad
n=1,2,\cdots,
$$
the matrix kernel of integral Fredholm equation is defined
by integral expression
$$
\hat J(k,k_1)=\dfrac{2}{\sqrt{\pi}}\int\limits_{0}^{\infty}
\dfrac{e^{-\mu^2}K(\mu)\mu d\mu}{(1+k^2\mu^2)(1+k_1^2\mu^2)}.
$$

It is obvious, that
$$
\hat J(k,0)=\hat T_1(k),\qquad \hat J(0,k_1)=\hat  T_1(k_1).
$$

\section{Solution of vector Fredholm equation}

For the solution of the equation (5.6) we will
search in the form of Neumann's polynoms
$$
E(k)=(2-q)g_T\Big[E_0(k)+E_1(k)q+E_2(k)q^2+\cdots+E_m(k)q^m\Big]
\eqno{(6.1)}
$$
$$
\left(\begin{array}{c}\varepsilon_n
\\\varepsilon_T\\\end{array}\right)=\dfrac{2-q}{q}g_T
\left(\begin{array}{c}\varepsilon_n^\circ+\varepsilon_n^1q+
\varepsilon_n^2q^2+\cdots+\varepsilon_n^mq^m
\\\varepsilon_T^\circ+\varepsilon_T^1q+\varepsilon_T^2q^2+\cdots+
\varepsilon_T^mq^m\\
\end{array}\right).
\eqno{(6.2)}
$$

Let us substitute (6.1) and (6.2) in the equation (5.6). We receive
system of the hooked equations
$$
L(k)E_0(k)=-\hat{T}_1(k)\left(\begin{array}{c}\varepsilon_n^\circ
\\\varepsilon_T^\circ\\\end{array}\right)+\hat{T}_2(k)
\left(\begin{array}{r}-1\\1\\\end{array}\right),
\eqno{(6.3)}
$$
$$
L(k)E_1(k)=-\hat{T}_1(k)\left(\begin{array}{c}\varepsilon_n^1
\\\varepsilon_T^1\\\end{array}\right)-\dfrac{1}{\pi}
\int\limits_{0}^{\infty}\hat J(k,k_1)E_0(k_1)dk_1,
\eqno{(6.4)}
$$
$$
...\qquad...\qquad...\qquad...\qquad...\qquad...\qquad...\qquad...
$$
$$
L(k)E_m(k)=-\hat{T}_1(k)\left(\begin{array}{c}\varepsilon_n^m
\\\varepsilon_T^m\\\end{array}\right)-\dfrac{1}{\pi}
\int\limits_{0}^{\infty}\hat J(k,k_1)E_{m-1}(k_1)dk_1,\quad m=1,2, \cdots
\eqno{(6.5)}
$$

Let us calculate in an explicit form the matrixes entering into
the equation (5.6)
$$\extrarowheight=10pt
L(k)=\left(\begin{array}{cc}1&0\\0&1\\\end{array}\right)-
\dfrac{2}{\sqrt{\pi}}\int\limits_{-\infty}^{\infty}
e^{-\mu^2}\left(
            \begin{array}{cc}
              1 & \mu^2-\dfrac{1}{2}\\
             \dfrac{2}{3}\Big(\mu^2-\dfrac{1}{2}\Big) & \dfrac{2}{3}
\Big(\mu^4-\mu^2+\dfrac{5}{4}\Big) \\
            \end{array}
          \right)\dfrac{d\mu}{1+k^2\mu^2}=
$$
$$\extrarowheight=16pt
=\left(
   \begin{array}{cc}
     1-T_0(k) & -T_2(k)+\dfrac{1}{2}T_0(k)\\
     -\dfrac{2}{3}\Big(T_2(k)-\dfrac{1}{2}T_0(k)\Big) & 1-\dfrac{2}{3}
\Big(T_4(k)-T_2(k)+\dfrac{5}{4}T_0(k)\Big) \\
   \end{array}
 \right).
$$

Here integrals are entered
$$
T_m(k)=\dfrac{2}{\sqrt{\pi}}\int\limits_{0}^{\infty}
\dfrac{e^{-\mu^2}\mu^m d\mu}{1+k^2\mu^2},\qquad m=0,1,2,\cdots .
$$

Let us notice, that the dispersion matrix $L(k)$ is proportional to $k^2$.
Let us notice, that
$$
\dfrac{2}{\sqrt{\pi}}\int\limits_{0}^{\infty}e^{-t^2}K(t)dt=
\dfrac{1}{\sqrt{\pi}}\int\limits_{-\infty}^{\infty}e^{-t^2}K(t)dt=E_2.
$$
Therefore the dispersion matrix--function is equal
$$
L(k)=E_2-\hat T_0(k)=
\dfrac{2}{\sqrt{\pi}}\int\limits_{0}^{\infty}e^{-t^2}K(t)dt-
\dfrac{2}{\sqrt{\pi}}\int\limits_{0}^{\infty}\dfrac{e^{-t^2}K(t)}
{1+k^2t^2}dt=
$$
$$
= k^2\dfrac{2}{\sqrt{\pi}}\int\limits_{0}^{\infty}
\dfrac{e^{-t^2}K(t)t^2dt}{1+k^2t^2}=k^2\hat T_2(k).
\eqno{(6.6)}
$$

Let us write out matrix elements $\hat T_2(k)$
$$
T^2_{11}(k)=T_2(k),\qquad
T^2_{12}(k)=\dfrac{2}{\sqrt{\pi}}
\int\limits_{0}^{\infty}\dfrac{e^{-t^2}(t^2-1/2)t^2dt}{1+k^2t^2}=
T_4(k)-\dfrac{1}{2}T_2(k),
$$
$$
T^2_{21}(k)=\dfrac{2}{3}T^2_{12}(k)=
\dfrac{2}{3}\Big(T_4(k)-\dfrac{1}{2}T_2(k)\Big),\quad
T^2_{22}(k)=\dfrac{2}{3}\Big(T_6(k)-T_4(k)+\dfrac{5}{4}T_2(k)\Big).
$$

Therefore
$$
\hat T_2(k)=\left(
              \begin{array}{cc}
                T_2(k) & T_4(k)-\dfrac{1}{2}T_2(k) \\
                \dfrac{2}{3}\Big(T_4(k)-\dfrac{1}{2}T_2(k)\Big) &
\dfrac{2}{3}\Big(T_6(k)-T_4(k)+\dfrac{5}{4}T_2(k)\Big) \\
              \end{array}
            \right).
$$

Further we find a matrix $\hat T_1(k)$
$$\extrarowheight=10pt
\hat T_1(k)=\left(
              \begin{array}{cc}
                T_1(k) & T_3(k)-\dfrac{1}{2}T_1(k)\\
\dfrac{2}{3}\Big(T_3(k)-\dfrac{1}{2}T_1(k)\Big) & \dfrac{2}{3}
\Big(T_5(k)-T_3(k)+\dfrac{5}{4}T_1(k)\Big) \\
              \end{array}
            \right).
$$

Let us start the solution of zero approximation. From the equation (6.3) with
by the help (6.6) it is received, that
$$
E_0(k)=\dfrac{1}{k^2}\hat T_2^{-1}(k)
\Bigg[-\hat{T}_1(k)\left(\begin{array}{c}\varepsilon_n^\circ
\\\varepsilon_T^\circ\\\end{array}\right)+\hat{T}_2(k)
\left(\begin{array}{r}-1\\1\\\end{array}\right)\Bigg]=
$$
$$
=-\dfrac{1}{k^2}\Big[\hat T_2^{-1}(k)\hat{T}_1(k)
\left(\begin{array}{c}\varepsilon_n^\circ
\\\varepsilon_T^\circ\\\end{array}\right)-
\left(\begin{array}{r}-1\\1\\\end{array}\right)\Big].
\eqno{(6.7)}
$$

For existence of the zero solution (6.7) we will eliminate at it
solution a pole of the second order.

Let us notice, that $T_m(k)$ it is possible to present integrals in
the form
$$
T_m(k)=T_m(0)-k^2T_{m+2}(k),
$$
where
$$
T_m(0)=\dfrac{2}{\sqrt{\pi}}\int\limits_{0}^{\infty}e^{-\mu^2}\mu^m
d\mu,\qquad m=0,1,2,\cdots .
$$

By means of these equalities we will transform matrixes
$\hat T_1(k)$ and $\hat
T_1(k)$
$$\extrarowheight=10pt
\hat T_1(k)=\dfrac{1}{\sqrt{\pi}}
\left(\begin{array}{cc}1 & \dfrac{1}{2} \\
\dfrac{1}{3} & \dfrac{3}{2} \\\end{array}\right)
$$
$$\extrarowheight=16pt
-k^2\left(\begin{array}{cc}T_3(k) &T_5(k)-\dfrac{1}{2}T_3(k)\\
\dfrac{2}{3}\Big(T_5(k)-\dfrac{1}{2}T_3(k)\Big)&
\dfrac{2}{3}\Big(T_7(k)-T_5(k)+\dfrac{5}{4}T_3(k)\Big)\\\end{array}\right)
$$
and
$$\extrarowheight=8pt
\hat T_2(k)=\dfrac{1}{2}
\left(\begin{array}{cc}1 &1\\
\dfrac{2}{3}& \dfrac{7}{3} \\\end{array}\right)
$$
$$\extrarowheight=16pt
-k^2\left(\begin{array}{cc}T_4(k) &T_6(k)-
\dfrac{1}{2}T_4(k)\\\dfrac{2}{3}\Big(T_6(k)-\dfrac{1}{2}T_4(k)\Big)&
\dfrac{2}{3}\Big(T_8(k)-T_6(k)+\dfrac{5}{4}T_4(k)\Big)\\\end{array}\right).
$$

Let us rewrite these equalities in the matrix form
$$
\hat T_1(k)=\hat T_1(0)-k^2\hat T_3(k), \qquad
\hat T_2(k)=\hat T_2(0)-k^2\hat T_4(k).
$$

Here
$$\extrarowheight=10pt
\hat T_1(0)=\dfrac{1}{\sqrt{\pi}}
\left(\begin{array}{cc}1 & \dfrac{1}{2} \\
\dfrac{1}{3} & \dfrac{3}{2} \\\end{array}\right),\qquad
\hat T_2(0)=\dfrac{1}{2}
\left(\begin{array}{cc}1 &1\\
\dfrac{2}{3}& \dfrac{7}{3} \\\end{array}\right),
$$
$$\extrarowheight=10pt
\hat T_3(k)=\left(\begin{array}{cc}T_3(k) &
T_5(k)-\dfrac{1}{2}T_3(k)\\
\dfrac{2}{3}\Big(T_5(k)-\dfrac{1}{2}T_3(k)\Big)&
\dfrac{2}{3}\Big(T_7(k)-T_5(k)+
\dfrac{5}{4}T_3(k)\Big)\\\end{array}\right),
$$
$$\extrarowheight=10pt
\hat T_4(k)=\left(\begin{array}{cc}T_4(k) &
T_6(k)-\dfrac{1}{2}T_4(k)\\
\dfrac{2}{3}\Big(T_6(k)-\dfrac{1}{2}T_4(k)\Big)&
\dfrac{2}{3}\Big(T_8(k)-T_6(k)+
\dfrac{5}{4}T_4(k)\Big)\\\end{array}\right).
$$

Let us return to equality (4.7) and by means of the previous equalities
let us write down it in the form
$$
E_0(k)=\dfrac{1}{k^2}\hat T_2^{-1}(k)\Bigg[-\hat T_1(0)
\left(\begin{array}{c}\varepsilon_n^\circ
\\\varepsilon_T^\circ\\\end{array}\right)+\hat T_2(0)
\left(\begin{array}{r}-1 \\1\\\end{array}\right)
\Bigg]+
$$
$$
+\hat T_2^{-1}(k)\Bigg[\hat T_3(k)\left(\begin{array}{c}
\varepsilon_n^\circ
\\\varepsilon_T^\circ\\\end{array}\right)-\hat T_4(k)
\left(\begin{array}{r}-1 \\1\\\end{array}\right)
\Bigg].
$$

For existence of zero approach we will demand, that
the following equality was carried out
$$
\hat T_1(0)
\left(\begin{array}{c}\varepsilon_n^\circ
\\\varepsilon_T^\circ\\\end{array}\right)=\hat T_2(0)
\left(\begin{array}{r}-1 \\1\\\end{array}\right).
\eqno{(6.8)}
$$

So, in zero approach taking into account (6.8) we receive, that
$$
\left(\begin{array}{c}\varepsilon_n
\\\varepsilon_T\\\end{array}\right)=\dfrac{2-q}{q}g_T
\left(\begin{array}{c}\varepsilon_n^\circ\\\varepsilon_T^\circ\\
\end{array}\right), \qquad
$$
and
$$
E(k)=(2-q)g_TE_0(k),
$$
where
$$
E_0(k)=\hat T_2^{-1}(k)\Bigg[\hat T_3(k)\left(\begin{array}{c}
\varepsilon_n^\circ
\\\varepsilon_T^\circ\\\end{array}\right)-\hat T_4(k)
\left(\begin{array}{r}-1 \\1\\\end{array}\right)
\Bigg]=
$$
$$
=\hat T_2^{-1}(k)\Bigg[\hat T_3(k)\hat T_1^{-1}(0)\hat T_2(0)-\hat T_4(k)
\Bigg]\left(\begin{array}{r}-1 \\1\\\end{array}\right).
\eqno{(6.9)}
$$

\section{The first and the higher approximations of solution}

From the equation (6.4) by means of (6.6) it is received the following equation
$$
k^2\hat T_2(k)E_1(k)=-\hat T_1(k)\left(\begin{array}{c}\varepsilon_n^1
\\\varepsilon_T^1\\\end{array}\right)-\dfrac{1}{\pi}
\int\limits_{0}^{\infty}\hat J(k,k_1)E_0(k_1)dk_1.
\eqno{(7.1)}
$$

Here the vector--column $E_0(k_1)$ is defined by expression (6.9).

Let us notice, that
$$
\hat J(k,k_1)=\hat T_1(k_1)-k^2\hat J_3(k,k_1).
$$

The equation (7.1) by means of this equality we will rewrite in the form
$$
k^2\hat T_2(k)E_1(k)=-\hat T_1(0)\left(\begin{array}{c}\varepsilon_n^1
\\\varepsilon_T^1\\\end{array}\right)-\dfrac{1}{\pi}
\int\limits_{0}^{\infty}\hat T_1(k_1)E_0(k_1)dk_1+
$$
$$
+k^2\Bigg[\hat T_3(k)
\left(\begin{array}{c}\varepsilon_n^1
\\\varepsilon_T^1\\\end{array}\right)+
\dfrac{1}{\pi}\int\limits_{0}^{\infty}\hat J_3(k,k_1)E_0(k_1)dk_1\Bigg].
\eqno{(7.2)}
$$

From the equation (7.2) it is visible, that for existence of the first
approximation we should impose on free parametres of the solution
the following vector condition
$$
\left(\begin{array}{c}\varepsilon_n^1
\\\varepsilon_T^1\\\end{array}\right)=-\hat T^{-1}_1(0)
\dfrac{1}{\pi}
\int\limits_{0}^{\infty}\hat T_1(k_1)E_0(k_1)dk_1.
\eqno{(7.3)}
$$

Then from the equation (7.2) it is found spectral density of our problem in
the first approximation
$$
E_1(k)=\hat T_2^{-1}(k)\Bigg[\hat T_3(k)
\left(\begin{array}{c}\varepsilon_n^1
\\\varepsilon_T^1\\\end{array}\right)+
\dfrac{1}{\pi}\int\limits_{0}^{\infty}\hat J_3(k,k_1)E_0(k_1)dk_1\Bigg].
\eqno{(7.4)}
$$

So, as  first approximation the problem solution looks as the following
the form
$$
E(k)=(2-q)g_T\Big[E_0(k)+E_1(k)q\Big],
$$
$$
\left(\begin{array}{c}\varepsilon_n
\\\varepsilon_T\\\end{array}\right)=\dfrac{2-q}{q}g_T
\left(\begin{array}{c}\varepsilon_n^\circ+\varepsilon_n^1q
\\\varepsilon_T^\circ+\varepsilon_T^1q\\\end{array}\right).
$$

Let us consider approximation of an arbitrary $m$th  order. From
the equations (7.5) by means of (6.6) we will write down
$$
k^2\hat T_2(k)E_m(k)=-\hat T_1(0)\left(\begin{array}{c}\varepsilon_n^m
\\\varepsilon_T^m\\\end{array}\right)-\dfrac{1}{\pi}
\int\limits_{0}^{\infty}\hat T_1(k_1)E_{m-1}(k_1)dk_1+
$$
$$
+k^2\Bigg[\hat T_3(k)
\left(\begin{array}{c}\varepsilon_n^m
\\\varepsilon_T^m\\\end{array}\right)+
\dfrac{1}{\pi}\int\limits_{0}^{\infty}
\hat J_3(k,k_1)E_{m-1}(k_1)dk_1\Bigg],\quad m=1,2,\cdots.
\eqno{(7.5)}
$$

For existence $m$-th approximation we will impose the
following condition on free parametres of solutions
$$
\left(\begin{array}{c}\varepsilon_n^m
\\\varepsilon_T^m\\\end{array}\right)=-\hat T^{-1}_1(0)
\dfrac{1}{\pi}
\int\limits_{0}^{\infty}\hat T_1(k_1)E_{m-1}(k_1)dk_1,\quad
m=1,2,\cdots.
\eqno{(7.6)}
$$

Now from the equation (7.5) it is found spectral density in
$m $-th approximation
$$
E_m(k)=\hat T_2^{-1}(k)\Bigg[\hat T_3(k)
\left(\begin{array}{c}\varepsilon_n^m
\\\varepsilon_T^m\\\end{array}\right)+
\dfrac{1}{\pi}\int\limits_{0}^{\infty}\hat J_3(k,k_1)E_{m-1}(k_1)dk_1\Bigg],
\quad m=1,2,\cdots.
\eqno{(7.5)}
$$

\begin{center}
  \item{}
\section{Numerical calculations and comparison with exact solution}
\end{center}

From the equation (6.8) it is found in zero approximation
free parametres of the solution
$$
\left(\begin{array}{c}\varepsilon_n^\circ
\\\varepsilon_T^\circ\\\end{array}\right)=\hat T^{-1}_1(0)\hat T_2(0)
\left(\begin{array}{r}-1 \\1\\\end{array}\right)=
\left(\begin{array}{r}-0.55389
\\1.10778\\\end{array}\right),
$$
or
$$
\varepsilon_n^\circ=-\dfrac{\varepsilon_T^\circ}{2}=
-\dfrac{5\sqrt{\pi}}{16}\approx -0.55389,\qquad
\varepsilon_T^\circ=\dfrac{5\sqrt{\pi}}{8}\approx 1.10778.
$$

For comparison we will give exact values of the dimensionless
coefficients of temperature  jump  and jump of numerical density
\cite{3} at diffusion reflexion of molecules from the wall
$\varepsilon_T=1.30272$ и $\varepsilon_n=-0.74428$.

Error in zero approximation for temperature jump
makes 15 \%, and for jump of numerical density makes 25.5 \%.

Let us consider the first approximation. From the equation (7.3) it is found
$$
\varepsilon_n^1=-\dfrac{3\sqrt{\pi}}{8}(3D_1-D_2),\qquad
\varepsilon_T^1=\dfrac{\sqrt{\pi}}{4}(D_1-3D_2),
\eqno{(8.1)}
$$
where $D_1$ and $D_2$ are defined by the following expression
$$
\left(\begin{array}{c}D_1\\D_2\\\end{array}\right)=\dfrac{1}{\pi}
\int\limits_{0}^{\infty}\hat T_1(k)E_0(k)dk.
\eqno{(8.3)}
$$

According to (6.9) for vector--column $E_0(k) $ we receive the expression
$$
E_0(k)=\hat T_2^{-1}(k)C(k),
\eqno{(8.4)}
$$
where
$$
C(k)=\hat T_3(k)\left(\begin{array}{c}
\varepsilon_n^\circ
\\\varepsilon_T^\circ\\\end{array}\right)-\hat T_4(k)
\left(\begin{array}{r}-1 \\1\\\end{array}\right).
$$

From here we find elements of vector--column $C(k)$
$$
C_1(k)=\varepsilon_T^\circ[T_5(k)-T_3(k)]+\dfrac{3}{2}T_4(k)-T_6(k),
$$
and
$$
C_2(k)=\dfrac{2}{3}\Big[\varepsilon_T^\circ\Big(T_7(k)-
\dfrac{3}{2}T_5(k)+\dfrac{3}{2}T_3(k)\Big)+2T_6(k)-T_8(k)-\dfrac{7}{4}
T_4(k)\Big].
$$

Now according to (8.4) we find  elements of the vector--column
$E_0(k)$
$$
E_0^1(k)=\dfrac{1}{\det \hat T_2(k)}\Big[\dfrac{2}{3}C_1(k)
(T_6(k)-T_4(k)+\dfrac{5}{4}T_2(k)-C_2(k)\Big(T_4(k)-
\dfrac{1}{2}T_2(k)\Big)\Big],
$$
and
$$
E_0^2(k)=\dfrac{1}{\det \hat T_2(k)}\Big[-\dfrac{2}{3}C_1(k)
(T_4(k)-\dfrac{1}{2}T_2(k))+C_2(k)T_2(k)\Big],
$$
where
$$
\det \hat
T_2(k)=\dfrac{2}{3}\Big[T_2(k)T_6(k)-T_4^2(k)+T_2^2(k)\Big].
$$

According to (8.3) it is found
$$
D_1=\dfrac{1}{\pi}\int\limits_{0}^{\infty}\Big[T_1(k)E_0^1(k)+
\Big(T_3(k)-\dfrac{1}{2}T_1(k)\Big)E_0^2(k)\Big]dk
$$
and
$$
D_2=\dfrac{1}{\pi}\int\limits_{0}^{\infty}
\Big[\dfrac{2}{3}\Big(T_3(k)-\dfrac{1}{2}T_1(k)\Big)E_0^1(k)+
\dfrac{2}{3}\Big(T_5(k)-T_3(k)+\dfrac{5}{4}T_1(k)\Big)E_0^2(k)\Big]dk.
$$

Under formulas (8.1) and (8.2) it is found as the first approximation, that
$\varepsilon_n^1=-0.21018$ и $\varepsilon_T^1=0.21378$.

Thus, as a first approximation we find, that
$$
\left(\begin{array}{c}\varepsilon_n
\\\varepsilon_T\\\end{array}\right)=\dfrac{2-q}{q}g_T
\left(\begin{array}{c}-0.55389-0.21018q
\\1.10778+0.21378q\\\end{array}\right).
$$

Comparison with exact result at $q=1$ shows, that
as the first approximation an error
in  finding of temperature jump  coefficient makes 1.4 \%,
and in finding of numerical density jump coefficient
makes 2.7 \%.

Let us spend comparison of the received results in
the present work with the results received in \cite{4}
by high-precision method of discrete ordinates.
Let us notice, that value of temperature jump coefficient at
$q=1$ from \cite{4} in accuracy to equally exact value (see \cite{3}, p.
228).

From given below table it is visible, that with
decreasing of  accommodation coefficient accuracy
of the first approximation grows and at $q=0.1$ the error makes
hundredth fraction of percent.

Table \bigskip
\begin{center}\normalsize
\begin{tabular}{|c|c|c|c|c|c|c|c|}
\hline
$q$        &1      &0.9    & 0.7   & 0.6   & 0.5   & 0.3   &0.1     \\\hline
$[4]$      &1.30272&1.57026&2.31753&2.86762&3.62922&6.63051&21.45012\\\hline
Present article &1.32156&1.58911&2.33522&2.88411&3.64401&6.64085&21.45400\\\hline
Error,\%  &-1.4\%    & -1.2\%&-0.75\%&-0.58\%& -0.44\%&-0.16\%& -0.018\%\\\hline
\end{tabular}

\end{center}

\begin{figure}[t]
\begin{center}
\includegraphics[width=14cm, height=12cm]{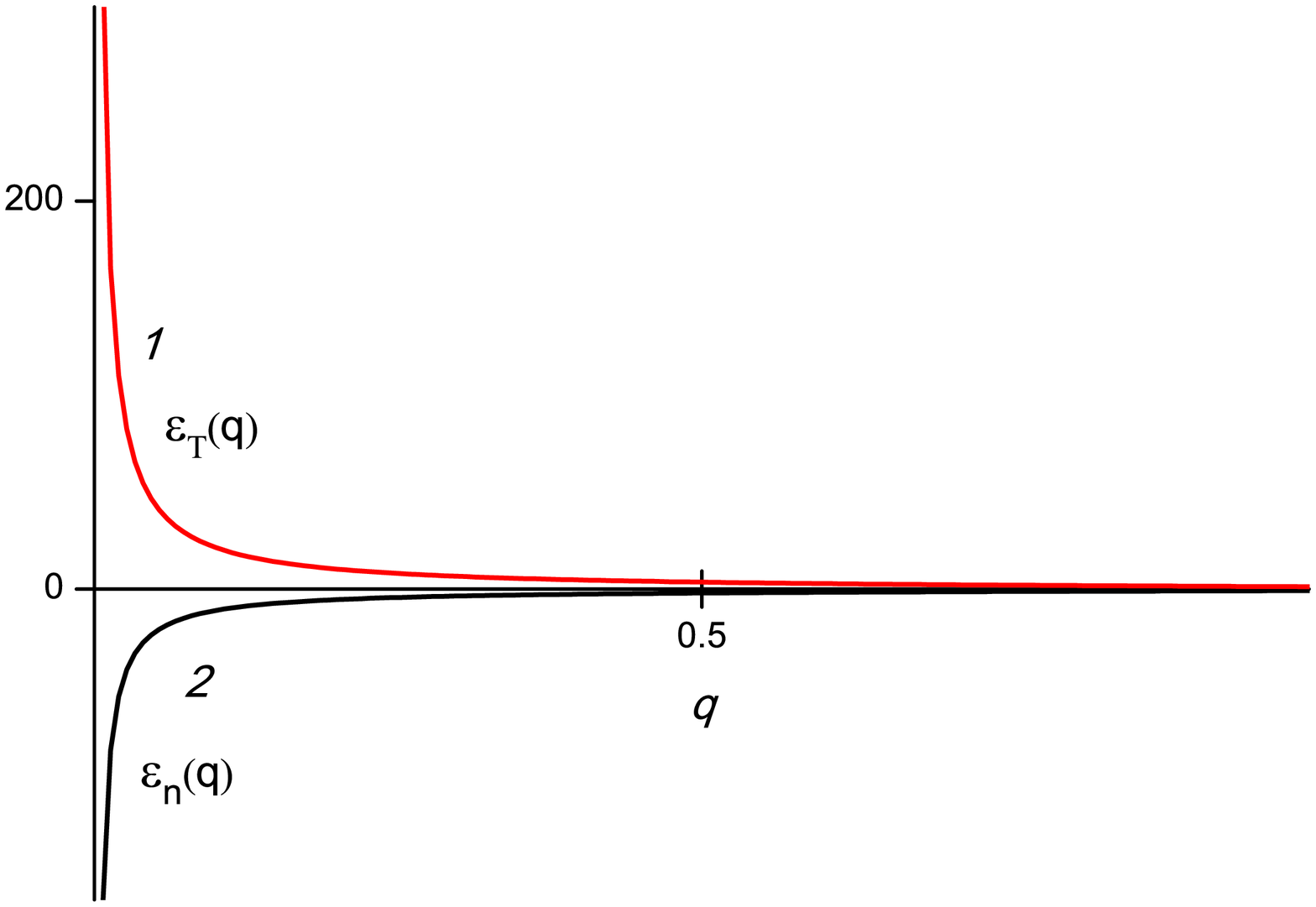}
\end{center}
\begin{center}
{Fig. Dependences $\varepsilon_T=\varepsilon_T(q)$ and
$\varepsilon_n=\varepsilon_n(q)$. The first approximation of solution.}
\end{center}
\end{figure}

\begin{center}
\item{}\section*{\bf Conclusion} 
\end{center}

In the present work the Smoluchowsky'  problem
about temperature jump with mirror--diffusion
boundary conditions is solved. The kinetic equation is used,
received as a linearization result of the modelling kinetic
Boltzmann equation in relaxation approximation (BGK--equation).
Then the problem is reduced to the solution of half-space
boundary problem for the vector kinetic equation with
matrix kernel.
The generalized method of a source develops.
This method has been offered in \cite{30}.
Comparison with well-known Barichello---Siewert'  high-exact results
is made. Zero and the first approach of jumps of temperature
and numerical density are received.
It is shown, that already the first approach leads
to the results close to the exact. Further in this direction
it is offered to use the models leading to correct Prandtl
number, for example, S-model of Shakhov  \cite{27}.

\end{document}